\documentclass[%
 reprint,
 amsmath,amssymb,
 aps,
pra,
]{revtex4-2}

\usepackage{graphicx}
\usepackage{dcolumn}
\usepackage{bm}
\usepackage{siunitx}
\usepackage{multirow}
\usepackage{diagbox}
\usepackage{comment}
\usepackage{xcolor}

\usepackage{xr}
\usepackage{xcolor}
\usepackage{url}

\begin{document}

\preprint{Mapping complex optical light field distributions with single fluorescent molecules}

\title{Mapping complex optical light field distributions with single fluorescent molecules}

\author{Daniel Marx}
\affiliation{Third Institute of Physics (Biophysics), Georg August University, 37077 Göttingen, Germany.}
\author{Ivan Gligonov}
\affiliation{Third Institute of Physics (Biophysics), Georg August University, 37077 Göttingen, Germany.}
\author{David Malsbenden}
\affiliation{Institute of Physical Chemistry, RWTH Aachen University, 52074 Aachen, Germany.}
\author{Dominik Wöll}
\affiliation{Institute of Physical Chemistry, RWTH Aachen University, 52074 Aachen, Germany.}
\author{Oleksii Nevskyi}
\affiliation{Third Institute of Physics (Biophysics), Georg August University, 37077 Göttingen, Germany.}
\author{Jörg Enderlein}%
 \homepage{https://www.joerg-enderlein.de}
 \email{jenderl@gwdg.de}
\affiliation{Third Institute of Physics (Biophysics), Georg August University, 37077 Göttingen, Germany.}
 \affiliation{Cluster of Excellence “Multiscale Bioimaging: from Molecular Machines to Networks of Excitable Cells” (MBExC), Universitätsmedizin Göttingen, Robert-Koch-Str. 40, 37077, Göttingen, Germany.}

\date{\today}


\begin{abstract}

\end{abstract}

\maketitle

\textbf{Introduction.} Fluorescent molecules behave as ideal point-like electric dipoles in both absorption and emission \cite{karedla2015simultaneous}. This fundamental property is the basis for several methods, such as fluorescence anisotropy spectroscopy \cite{lakowicz1999fluorescence} and polarization-resolved fluorescence correlation spectroscopy \cite{ehrenberg1976fluorescence,pieper2011fluorescence}, which measure molecular rotational diffusion. The dipolar nature of single-molecule emission has been studied extensively through various techniques, including defocused imaging \cite{dickson1998simultaneous,bartko1999imaging,Enderlein2003,karedla2015simultaneous}, back focal plane imaging \cite{lieb2004single,backer2014extending}, complex polarization-resolved detection \cite{beausang2013tilting,backer2016enhanced,ding2020single,rimoli20224polar,zhang2023six}, and polarization point spread function engineering \cite{curcio2020birefringent,ding2021single,hulleman2021simultaneous}.

The dipolar nature of single-molecule absorption has been investigated with techniques like excitation polarization modulation \cite{empedocles1999three,empedocles1999photoluminescence,ma2018anisotropic} or a combination of azimuthally and radially polarized excitation beams \cite{chizhik2011excitation,karedla2015simultaneous,ghosh2019excitation,zhang2022resolving}. These methods are widely employed for determining single-emitter absorption dipole orientations. In a more subtle application, single fluorescent molecules, acting as point-like dipole absorbers, have been exploited to map the electric field distribution in complex light fields \cite{rotenberg2014mapping} with nanometer-scale resolution \cite{cang2011probing,singh2014vectorial,steuwe2015visualizing, mack2017decoupling,novak2025using}.

Due to the principle of reciprocity, a $z$-stack of fluorescence images from a single, fixed molecule closely resembles a three-dimensional scan image recorded with a confocal laser scanning microscope (CLSM) using linearly polarized excitation \cite{karedla2015simultaneous}. However, this situation changes dramatically with a more complex polarization structure, such as circularly polarized light. In this manuscript, we use individual fluorescent molecules with fixed dipole orientations as probes to map the polarization and vectorial structure of non-trivial focused light fields, specifically tightly focused laser beams with circular versus linear polarization. We model the interaction using a full vectorial wave-optical description of the optical field. For our experiments, we use single terrylene diimide molecules embedded in a rigid polymer matrix. By systematically varying the excitation polarization (linear and both left- and right-handed circular), we directly reveal the chiral and vectorial nature of the focused light fields. Our theoretical models quantitatively reproduce the resulting three-dimensional excitation patterns, which allows for the precise extraction of both the molecules' absolute dipole orientations and the field characteristics.

\textbf{Theoretical background.} Following the seminal works of Wolf \cite{wolf1959electromagnetic} and Richards and Wolf \cite{richards1959electromagnetic} on light focusing and imaging through high numerical aperture optical systems, we describe the electric field distribution within the focus of a confocal laser scanning microscope as a superposition of electromagnetic plane waves. For a plane wave focused by a microscope objective, this superposition is given by:
\begin{equation}
\begin{split}
\mathbf{E}(\mathbf{r}) \propto &\int_0^{\Theta} d\theta \sin\theta \sqrt{\cos\theta} \int_0^{2\pi} d\phi \\
&\Bigg. \left[T_\parallel(\mathbf{k}) E_{0,\parallel}(\rho,\phi) \hat{\mathbf{e}}_\parallel + T_\perp(\mathbf{k}) E_{0,\perp}(\rho,\phi) \hat{\mathbf{e}}'_\perp\right] \\ &\quad\quad\exp\left[i\mathbf{k}\cdot\mathbf{r}+\Phi(\theta,\phi)\right],
\end{split}
\label{eq:Efocus}
\end{equation}
where $\mathbf{k}=2\pi n/\lambda\left(\cos\phi \sin\theta, \sin\phi \sin\theta, \cos\theta\right)$ is the wave vector of a plane wave with wavelength $\lambda$ (the excitation light wavelength). The electric field of the incoming plane wave in the back focal plane is expanded into its azimuthally ($E_{0,\parallel}$) and radially ($E_{0,\perp}$) polarized components. The radial coordinate $\rho$ in the back focal plane is related to the propagation angle $\theta$ of a plane wave in the sample space by the relation $\rho = n \sin\theta$, where $n$ is the refractive index of the sample medium. The factor $\sqrt{\cos\theta}$ accounts for energy conservation when focusing a plane wave through a high-aperture objective. The unit vectors $\hat{\mathbf{e}}_\parallel$ and $\hat{\mathbf{e}}'_\perp$ are perpendicular to the wave vector $\mathbf{k}$ and correspond to the polarizations of the azimuthal and radial components \emph{after} focusing through the objective and traversing any intermediate parallel layers with different refractive indices placed between the objective lens and the sample. $T_\parallel(\mathbf{k})$ and $T_\perp(\mathbf{k})$ are the corresponding Fresnel transmission coefficients for $s$- and $p$-waves traveling along $\mathbf{k}$ in the sample space. 
The function $\Phi(\theta,\phi)$ accounts for potential phase distortions due to optical aberrations of the system. 
The upper integration limit $\Theta$ is the maximum half-angle of light collection of the objective, which determines its numerical aperture (N.A.) via $\mathrm{N.A.}=n \sin\Theta$. A more detailed explanation can be found in a recent review by \citeauthor{fazel2024fluorescence} \cite{fazel2024fluorescence}.

Different excitation polarization modes are described by an appropriate choice of the field amplitudes $E_{0,\parallel}(\phi)$ and $E_{0,\perp}(\phi)$. For an incoming plane wave linearly polarized along $\phi=0$, one has $E_{0,\parallel}\propto-\sin\phi$ and $E_{0,\perp}\propto\cos\phi$. A polarization along $\phi=\pi/2$ corresponds to $E_{0,\parallel}\propto\cos\phi$ and $E_{0,\perp}\propto\sin\phi$. A circularly polarized beam is described by $E_{0,\parallel}\propto -\sin\phi \pm i \cos\phi$ and $E_{0,\perp}\propto\cos\phi \pm i \sin\phi$, where the plus sign denotes right-handed circular polarization and the minus sign denotes left-handed circular polarization. The integration over the angle $\phi$ in eq.~\eqref{eq:Efocus} can be performed using Bessel's integral:
\begin{equation}
\begin{aligned}
\int_0^{2\pi} d\phi e^{in\phi - i\xi\cos(\phi-\psi)}& = 2\pi i^ne^{in\psi} J_n(\xi),
\end{aligned}
\end{equation}
while the integration over $\theta$ is carried out numerically using a Fast Fourier Transform \cite{leutenegger2006fast}.

Finally, the excitation efficiency of a single molecule with an electric dipole moment $\mathbf{p}$ is given by $\left\vert\mathbf{E}(\mathbf{r}_0-\mathbf{r}_f)\cdot\mathbf{p}(\mathbf{r}_0)\right\vert^2$, where $\mathbf{r}_f$ is the center position of the scanning focus and $\mathbf{r}_0$ is the fixed position of the molecule in the sample. This value is proportional to the detectable fluorescence signal and, as a function of the focus position $\mathbf{r}_f$, yields the observable scan image of a single molecule. Note that both $\mathbf{r}_f$ and $\mathbf{r}_0$ are three-dimensional vectors, so our derivations also apply to three-dimensional scan images where the excitation focus is scanned not only laterally but also axially.

\begin{figure}[!h]
\centering
\includegraphics[width=\linewidth]{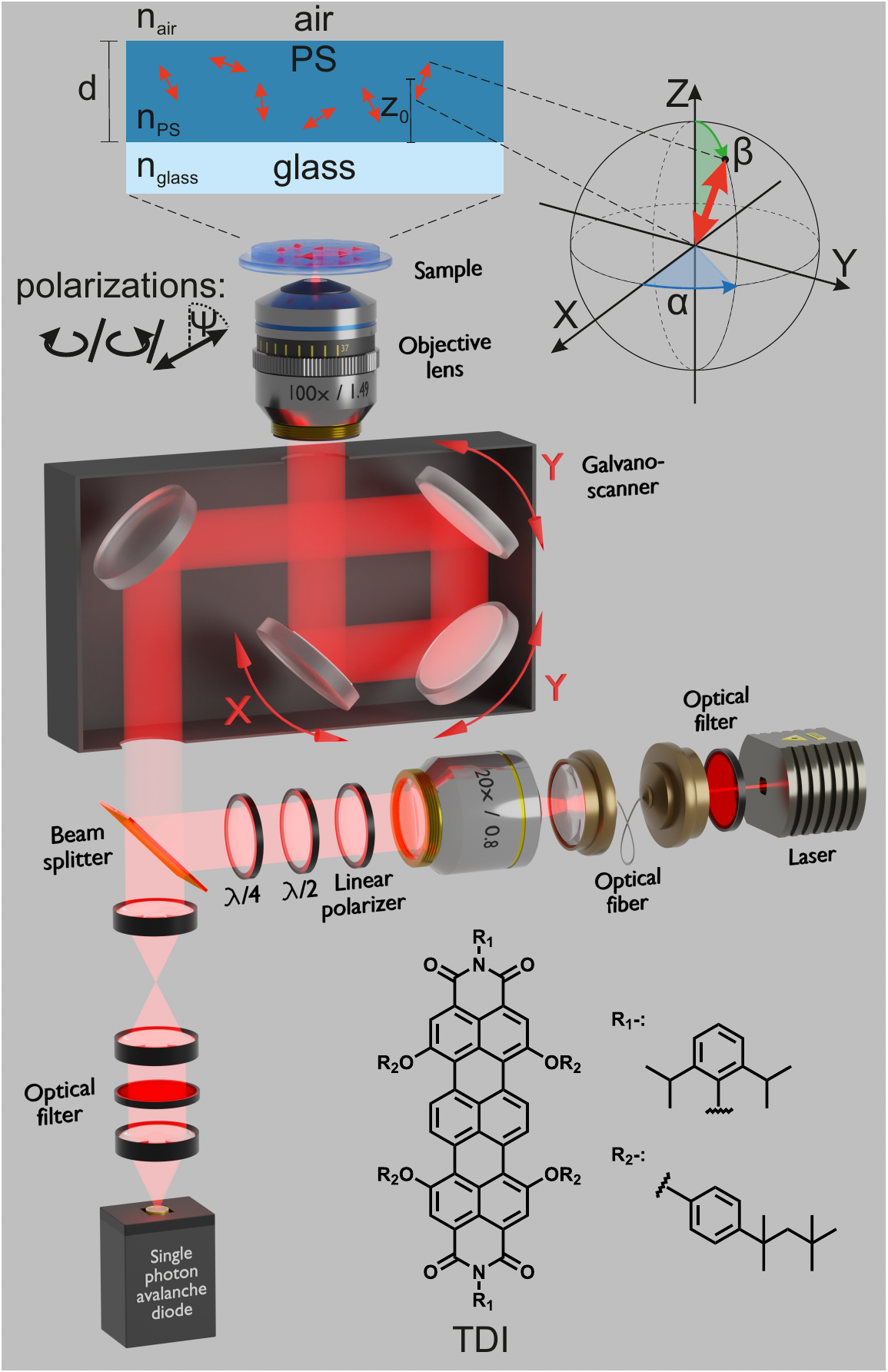}
\caption{Schematic of the custom-built confocal laser scanning microscope used for single-molecule experiments. The microscope is optimized for imaging individual terrylene diimide (TDI) molecules embedded in a thin (30 nm) polystyrene (PS) film. In contrast to a standard confocal setup, the pinhole is removed to maximize photon collection efficiency. Polarization control is achieved by rotating the $\lambda/2$ and $\lambda/4$-plates, enabling left- and right-handed circular polarization as well as linear polarization with an adjustable orientation $\Psi$.
On top is a sketch of the sample defining the corresponding parameters. A detailed description of the setup and sample preparation is provided in the Supplemental Material Sections A and B.}
\label{fig:setup}
\end{figure}

\textbf{Results.} To investigate the polarization-dependent excitation patterns of individual molecules, we used a highly photostable derivative of terrylene diimide (TDI, see Figure \ref{fig:setup}) embedded in a thin polystyrene (PS, $T_g  > 90~^\circ$C) film with a thickness of $d = 30~\mathrm{nm}$ (see Supplemental Material Section D1). At room temperature, the polymer matrix is sufficiently rigid to immobilize the dyes in a fixed orientation, and the refractive index of thin PS films is well known \cite{james2019surface,vignaud2014densification}. Measurements were performed on a custom-built CLSM shown in Figure \ref{fig:setup}, and $z$-stacks of images were recorded with a 50~nm distance between focal planes.

\begin{figure*}[t]
\centering
\includegraphics[width=0.7\textwidth]{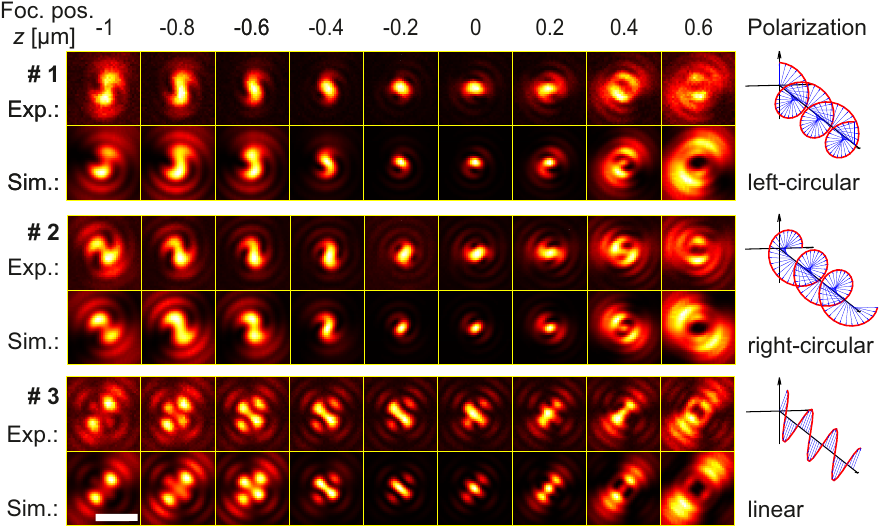}
\caption{Experimental and corresponding theoretical patterns for three TDI molecules with fixed orientations embedded in a thin PS film. For each molecule, images were recorded at different focal positions $z$ of the objective.
Negative $z$-values correspond to moving the objective closer to the sample. The first molecule was measured using left-handed circular polarization, the second with right-handed circular polarization, and the third with linear polarization oriented at $\Psi=82^\circ$.
All images are normalized to their respective maximum intensity. The exact parameters used for calculating the theoretical patterns are provided in the Supplemental Material Section D. The scale bar is $1~$\textmu m.
}
\label{fig:zstacks}
\end{figure*}

Owing to the exceptional photostability of TDI, it was possible to record the excitation pattern of a single molecule over a series of axial positions $z$ of the laser focus. Representative results for three different molecules are shown in Figure \ref{fig:zstacks}: one measured with left-handed circular polarization, one with right-handed circular polarization, and one with linear polarization. The corresponding theoretical patterns were calculated using the model described above. In this case, only primary spherical aberrations are considered (for more detail, see Supplemental Material Section D2).

\begin{figure*}[t]
\centering
\includegraphics[width=0.7\textwidth]{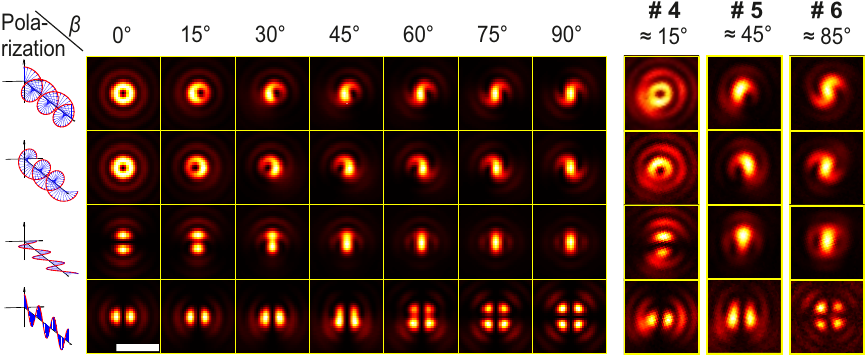}
\caption{Left: Theoretical patterns at the focal position $z=-0.35~$\textmu m for different out-of-plane angles $\beta$ and polarizations (from top to bottom: left-handed circular, right-handed circular, linear $\Psi =0^\circ $ and linear $\Psi=90^\circ$). Details of the parameters used for the calculations are provided in the Supplemental Information.
Right: Experimental patterns of three different TDI molecules inside a PS film. They are measured at $z=-0.35~$\textmu m with left- and right-handed circular, and two linear polarizations (from top to bottom). The directions of the linear polarizations $\Psi$ were chosen such that $\Psi=\alpha$ for the second images from the bottom and $\Psi=\alpha+90^\circ$ for the bottom images. For comparison, the experimental patterns are rotated by their estimated angle $\alpha$ (see Supplemental Material, Table S2). The scale bar is $1~$\textmu m.}
\label{fig:patterns}
\end{figure*}

The molecular orientations and positions were determined by a pattern-matching algorithm based on two-dimensional cross-correlation (see Supplemental Material Section C) \cite{lewis1995fast}. In this approach, the unknown parameters are the in-plane angle $\alpha$, the out-of-plane angle $\beta$, and the lateral position of the molecule inside the polymer film.
All other parameters were known and their values were chosen according to the properties of the optical setup and the sample (see Supplemental Material Section D). For the cross-correlation analysis, we first calculate theoretical images for discrete values of $\alpha$ and $\beta$ spaced $1^\circ$ apart. Cross-correlating these images with the experimental ones then allows for the determination of the molecule's lateral position and orientation. Using this approach, the orientation angles for the three molecules presented in Figure~\ref{fig:zstacks} were estimated to be $\alpha_{\#1}=227^\circ$, $\beta_{\#1}=61^\circ$; $\alpha_{\#2}=331^\circ$, $\beta_{\#2}=75^\circ$; and $\alpha_{\#3}=197^\circ$, $\beta_{\#3}=85^\circ$ (see Supplemental Material, Table S2).

While the linear polarization images exhibit the mirror symmetry typically seen in defocused wide-field images of electric dipole emitters, the images obtained with circular polarization exhibit a broken mirror symmetry and reflect the chirality of the excitation polarization. This peculiar chiral structure of three-dimensional scan images of single molecules is corroborated by the theoretical model calculations also shown in Figure~\ref{fig:zstacks}. As can be seen, the agreement between theoretical patterns and experimental data is excellent when the molecules are close to the focal plane (in focus) but slightly deteriorates further away from this plane. We attribute this to minor, unconsidered aberrations of the imaging system.


Although the dye exhibits high photostability, it is still limited in the context of the prolonged measurements required to acquire all three full $z$-stacks from a single molecule without bleaching. Therefore, the stacks for the three different excitation polarizations shown in Figure~\ref{fig:zstacks} were recorded on three separate molecules.

We also recorded two-dimensional scan images of a single molecule with a fixed focal plane but with four different polarization modalities: left- and right-handed circular polarization, as well as horizontal (along $\phi=0$) and vertical (along $\phi=\pi/2$) linear polarization. During all measurements, a custom-built autofocus system described by \citeauthor{radmacher2025fluorescence} \cite{radmacher2025fluorescence} maintained a stable focal position of $z=-0.35~$\textmu m. The linear polarizations were chosen relative to the in-plane orientation of the individual molecules, which had to be determined beforehand by rotating the linear polarization direction until it matched the in-plane orientation of a molecule. The result is shown in Figure~\ref{fig:patterns} for three different molecules, alongside theoretical model calculations. For ease of visual comparison, the experimental patterns in the figure were rotated by their estimated in-plane orientation angles $\alpha$: $\alpha_{\#4}=72^\circ$, $\alpha_{\#5}=196^\circ$, and $\alpha_{\#6}=252^\circ$. The out-of-plane angles $\beta$ were estimated using pattern matching, yielding $\beta_{\#4}=15^\circ$, $\beta_{\#5}=45^\circ$, and $\beta_{\#6}=85^\circ$. Once again, we find excellent agreement between experiment and theory, and the results clearly demonstrate the chiral and non-chiral structures of the single-molecule scan images, corresponding directly to the chirality of the excitation polarization.       

\textbf{Conclusion.} We investigated the interaction between light and single molecules by measuring their excitation patterns using a confocal microscope. We used three different polarization states: left-handed circular, right-handed circular, and linear. Furthermore, we developed a theoretical model to predict what these patterns should look like.

Our most significant finding is that \emph{circular polarization} creates highly \emph{chiral image patterns}. These patterns are strikingly different from the mirror-symmetric images typically seen with linear polarization or in defocused wide-field images. While one might naively expect that the fast rotation of circular polarization would average out the image to a simple, symmetrical pattern, our results prove the opposite. The scan images of these dipolar emitters can indeed reveal the intrinsic chiral nature of the excitation light \cite{bliokh2015spin}. This suggests that single molecules could serve as powerful probes for studying the topological structure of even more complex light fields.

Achieving close agreement between our experimental data and theoretical models required an exceptionally accurate understanding of numerous parameters for both the molecular probe and the optical setup. Even minor deviations in these parameters can lead to substantial differences in the predicted patterns. Nonetheless, our results demonstrate that with a careful and precise determination of these relevant parameters, we can achieve excellent agreement between the theoretical and experimental patterns. This was confirmed for multiple polarization states, across a wide range of focal positions ($z$), and for various molecular orientations (see Figures \ref{fig:zstacks} and \ref{fig:patterns}). We observed only minor discrepancies, which can likely be attributed to local variations in the sample (e.g., polymer film thickness or precise molecular position), refractive indices, numerical aperture, or optical aberrations not accounted for in our model.

The remarkable precision of our theoretical framework is crucial because the pattern-matching algorithm relies on these accurate predictions to reliably determine the absolute orientations of single molecules. Furthermore, knowing the expected patterns allows us to correct for small localization errors that can arise from asymmetric or shifted intensity distributions. Since the polarization states we considered—linear and circular—can be generated simply by inserting standard $\lambda$/2- and $\lambda$/4-wave plates, our approach is compatible with many existing confocal microscope setups.

Our method could also be used to estimate laser polarization by matching theoretical and experimental patterns. For example, patterns from horizontally oriented molecules can reveal the chirality of the laser's circular polarization, while patterns from vertically oriented molecules can be used to determine the linear polarization angle, as these are independent of the in-plane molecular angle. The achievable precision of such polarization measurements requires further systematic evaluation.

Beyond its demonstrated ability to predict single-molecule excitation patterns with high accuracy, our framework has broad potential across multiple disciplines. In biophysics, it could be applied to determine the orientation and localization of fluorophores in complex biological assemblies, aiding studies of protein organization, membrane dynamics \cite{borner2016single}, and cytoskeletal architecture. The method can also be used to probe molecular alignment in self-assembly systems \cite{adamczyk2022dna}, as well as to characterize optical anisotropy and nanoscale heterogeneity in materials such as polymers \cite{pressly2019increased}, liquid crystals, and nanocomposites.

Its compatibility with many existing confocal microscopes makes it a widely accessible tool for quantifying light-matter interactions without major hardware modifications. Our method could also serve as a diagnostic tool for characterizing and calibrating polarization states in optical systems or for correcting polarization-induced localization errors in super-resolution microscopy. By combining polarization-resolved excitation with pattern-matching analysis, this framework enables studies of rotational diffusion and orientational dynamics at the single-molecule level and could be extended to other imaging modalities, such as light-sheet or multiphoton microscopy, for three-dimensional orientation mapping in complex samples.
                  
\begin{acknowledgments}

Daniel Marx, David Malsbenden, Dominik W\"oll and J\"org Enderlein acknowledge financial support by the Deutsche Forschungsgemeinschaft (DFG, German Research Foundation) through project n$^\circ$ 511200316. 
Ivan Gligonov acknowledges funding by the International Max Planck Research School for Physics of Biological and Complex Systems and by the European Union via the HORIZON–MSCA–2022–DN ``Improving BiomEdical diagnosis through LIGHT-based technologies and machine learning -- BE-LIGHT'' (GA n$^\circ$101119924 -- BE-LIGHT). 
J\"org Enderlein and Oleksii Nevskyi acknowledge financial support by the European Research Council (ERC) for financial support via project ``smMIET'' (grant agreement n$^\circ$884488) under the European Union's Horizon 2020 research and innovation program. 
J\"org Enderlein acknowledges financial support by the Deutsche Forschungsgemeinschaft (DFG, German Research Foundation) through Germany’s Excellence Strategy EXC 2067/1-390729940. 
\end{acknowledgments}


\begin{thebibliography}{41}%
\makeatletter
\providecommand \@ifxundefined [1]{%
 \@ifx{#1\undefined}
}%
\providecommand \@ifnum [1]{%
 \ifnum #1\expandafter \@firstoftwo
 \else \expandafter \@secondoftwo
 \fi
}%
\providecommand \@ifx [1]{%
 \ifx #1\expandafter \@firstoftwo
 \else \expandafter \@secondoftwo
 \fi
}%
\providecommand \natexlab [1]{#1}%
\providecommand \enquote  [1]{``#1''}%
\providecommand \bibnamefont  [1]{#1}%
\providecommand \bibfnamefont [1]{#1}%
\providecommand \citenamefont [1]{#1}%
\providecommand \href@noop [0]{\@secondoftwo}%
\providecommand \href [0]{\begingroup \@sanitize@url \@href}%
\providecommand \@href[1]{\@@startlink{#1}\@@href}%
\providecommand \@@href[1]{\endgroup#1\@@endlink}%
\providecommand \@sanitize@url [0]{\catcode `\\12\catcode `\$12\catcode `\&12\catcode `\#12\catcode `\^12\catcode `\_12\catcode `\%12\relax}%
\providecommand \@@startlink[1]{}%
\providecommand \@@endlink[0]{}%
\providecommand \url  [0]{\begingroup\@sanitize@url \@url }%
\providecommand \@url [1]{\endgroup\@href {#1}{\urlprefix }}%
\providecommand \urlprefix  [0]{URL }%
\providecommand \Eprint [0]{\href }%
\providecommand \doibase [0]{http://dx.doi.org/}%
\providecommand \selectlanguage [0]{\@gobble}%
\providecommand \bibinfo  [0]{\@secondoftwo}%
\providecommand \bibfield  [0]{\@secondoftwo}%
\providecommand \translation [1]{[#1]}%
\providecommand \BibitemOpen [0]{}%
\providecommand \bibitemStop [0]{}%
\providecommand \bibitemNoStop [0]{.\EOS\space}%
\providecommand \EOS [0]{\spacefactor3000\relax}%
\providecommand \BibitemShut  [1]{\csname bibitem#1\endcsname}%
\let\auto@bib@innerbib\@empty
\bibitem [{\citenamefont {Karedla}\ \emph {et~al.}(2015)\citenamefont {Karedla}, \citenamefont {Stein}, \citenamefont {H{\"a}hnel}, \citenamefont {Gregor}, \citenamefont {Chizhik},\ and\ \citenamefont {Enderlein}}]{karedla2015simultaneous}%
  \BibitemOpen
  \bibfield  {author} {\bibinfo {author} {\bibfnamefont {N.}~\bibnamefont {Karedla}}, \bibinfo {author} {\bibfnamefont {S.~C.}\ \bibnamefont {Stein}}, \bibinfo {author} {\bibfnamefont {D.}~\bibnamefont {H{\"a}hnel}}, \bibinfo {author} {\bibfnamefont {I.}~\bibnamefont {Gregor}}, \bibinfo {author} {\bibfnamefont {A.}~\bibnamefont {Chizhik}}, \ and\ \bibinfo {author} {\bibfnamefont {J.}~\bibnamefont {Enderlein}},\ }\href@noop {} {\bibfield  {journal} {\bibinfo  {journal} {Phys. Rev. Lett.}\ }\textbf {\bibinfo {volume} {115}},\ \bibinfo {pages} {173002} (\bibinfo {year} {2015})}\BibitemShut {NoStop}%
\bibitem [{\citenamefont {Lakowicz}(1999)}]{lakowicz1999fluorescence}%
  \BibitemOpen
  \bibfield  {author} {\bibinfo {author} {\bibfnamefont {J.~R.}\ \bibnamefont {Lakowicz}},\ }in\ \href@noop {} {\emph {\bibinfo {booktitle} {Principles of fluorescence spectroscopy}}}\ (\bibinfo  {publisher} {Springer},\ \bibinfo {year} {1999})\ pp.\ \bibinfo {pages} {291--319}\BibitemShut {NoStop}%
\bibitem [{\citenamefont {Ehrenberg}\ and\ \citenamefont {Rigler}(1976)}]{ehrenberg1976fluorescence}%
  \BibitemOpen
  \bibfield  {author} {\bibinfo {author} {\bibfnamefont {M.}~\bibnamefont {Ehrenberg}}\ and\ \bibinfo {author} {\bibfnamefont {R.}~\bibnamefont {Rigler}},\ }\href@noop {} {\bibfield  {journal} {\bibinfo  {journal} {Q. Rev. Biophys.}\ }\textbf {\bibinfo {volume} {9}},\ \bibinfo {pages} {69} (\bibinfo {year} {1976})}\BibitemShut {NoStop}%
\bibitem [{\citenamefont {Pieper}\ and\ \citenamefont {Enderlein}(2011)}]{pieper2011fluorescence}%
  \BibitemOpen
  \bibfield  {author} {\bibinfo {author} {\bibfnamefont {C.~M.}\ \bibnamefont {Pieper}}\ and\ \bibinfo {author} {\bibfnamefont {J.}~\bibnamefont {Enderlein}},\ }\href@noop {} {\bibfield  {journal} {\bibinfo  {journal} {Chem. Phys. Lett.}\ }\textbf {\bibinfo {volume} {516}},\ \bibinfo {pages} {1} (\bibinfo {year} {2011})}\BibitemShut {NoStop}%
\bibitem [{\citenamefont {Dickson}\ \emph {et~al.}(1998)\citenamefont {Dickson}, \citenamefont {Norris},\ and\ \citenamefont {Moerner}}]{dickson1998simultaneous}%
  \BibitemOpen
  \bibfield  {author} {\bibinfo {author} {\bibfnamefont {R.~M.}\ \bibnamefont {Dickson}}, \bibinfo {author} {\bibfnamefont {D.~J.}\ \bibnamefont {Norris}}, \ and\ \bibinfo {author} {\bibfnamefont {W.}~\bibnamefont {Moerner}},\ }\href@noop {} {\bibfield  {journal} {\bibinfo  {journal} {Phys. Rev. Lett.}\ }\textbf {\bibinfo {volume} {81}},\ \bibinfo {pages} {5322} (\bibinfo {year} {1998})}\BibitemShut {NoStop}%
\bibitem [{\citenamefont {Bartko}\ and\ \citenamefont {Dickson}(1999)}]{bartko1999imaging}%
  \BibitemOpen
  \bibfield  {author} {\bibinfo {author} {\bibfnamefont {A.~P.}\ \bibnamefont {Bartko}}\ and\ \bibinfo {author} {\bibfnamefont {R.~M.}\ \bibnamefont {Dickson}},\ }\href@noop {} {\bibfield  {journal} {\bibinfo  {journal} {J. Phys. Chem. B}\ }\textbf {\bibinfo {volume} {103}},\ \bibinfo {pages} {11237} (\bibinfo {year} {1999})}\BibitemShut {NoStop}%
\bibitem [{\citenamefont {B\"{o}hmer}\ and\ \citenamefont {Enderlein}(2003)}]{Enderlein2003}%
  \BibitemOpen
  \bibfield  {author} {\bibinfo {author} {\bibfnamefont {M.}~\bibnamefont {B\"{o}hmer}}\ and\ \bibinfo {author} {\bibfnamefont {J.}~\bibnamefont {Enderlein}},\ }\href {\doibase 10.1364/JOSAB.20.000554} {\bibfield  {journal} {\bibinfo  {journal} {J. Opt. Soc. Am. B}\ }\textbf {\bibinfo {volume} {20}},\ \bibinfo {pages} {554} (\bibinfo {year} {2003})}\BibitemShut {NoStop}%
\bibitem [{\citenamefont {Lieb}\ \emph {et~al.}(2004)\citenamefont {Lieb}, \citenamefont {Zavislan},\ and\ \citenamefont {Novotny}}]{lieb2004single}%
  \BibitemOpen
  \bibfield  {author} {\bibinfo {author} {\bibfnamefont {M.~A.}\ \bibnamefont {Lieb}}, \bibinfo {author} {\bibfnamefont {J.~M.}\ \bibnamefont {Zavislan}}, \ and\ \bibinfo {author} {\bibfnamefont {L.}~\bibnamefont {Novotny}},\ }\href@noop {} {\bibfield  {journal} {\bibinfo  {journal} {J. Opt. Soc. Am. B}\ }\textbf {\bibinfo {volume} {21}},\ \bibinfo {pages} {1210} (\bibinfo {year} {2004})}\BibitemShut {NoStop}%
\bibitem [{\citenamefont {Backer}\ and\ \citenamefont {Moerner}(2014)}]{backer2014extending}%
  \BibitemOpen
  \bibfield  {author} {\bibinfo {author} {\bibfnamefont {A.~S.}\ \bibnamefont {Backer}}\ and\ \bibinfo {author} {\bibfnamefont {W.}~\bibnamefont {Moerner}},\ }\href@noop {} {\bibfield  {journal} {\bibinfo  {journal} {J. Phys. Chem. B}\ }\textbf {\bibinfo {volume} {118}},\ \bibinfo {pages} {8313} (\bibinfo {year} {2014})}\BibitemShut {NoStop}%
\bibitem [{\citenamefont {Beausang}\ \emph {et~al.}(2013)\citenamefont {Beausang}, \citenamefont {Shroder}, \citenamefont {Nelson},\ and\ \citenamefont {Goldman}}]{beausang2013tilting}%
  \BibitemOpen
  \bibfield  {author} {\bibinfo {author} {\bibfnamefont {J.~F.}\ \bibnamefont {Beausang}}, \bibinfo {author} {\bibfnamefont {D.~Y.}\ \bibnamefont {Shroder}}, \bibinfo {author} {\bibfnamefont {P.~C.}\ \bibnamefont {Nelson}}, \ and\ \bibinfo {author} {\bibfnamefont {Y.~E.}\ \bibnamefont {Goldman}},\ }\href@noop {} {\bibfield  {journal} {\bibinfo  {journal} {Biophys. J.}\ }\textbf {\bibinfo {volume} {104}},\ \bibinfo {pages} {1263} (\bibinfo {year} {2013})}\BibitemShut {NoStop}%
\bibitem [{\citenamefont {Backer}\ \emph {et~al.}(2016)\citenamefont {Backer}, \citenamefont {Lee},\ and\ \citenamefont {Moerner}}]{backer2016enhanced}%
  \BibitemOpen
  \bibfield  {author} {\bibinfo {author} {\bibfnamefont {A.~S.}\ \bibnamefont {Backer}}, \bibinfo {author} {\bibfnamefont {M.~Y.}\ \bibnamefont {Lee}}, \ and\ \bibinfo {author} {\bibfnamefont {W.}~\bibnamefont {Moerner}},\ }\href@noop {} {\bibfield  {journal} {\bibinfo  {journal} {Optica}\ }\textbf {\bibinfo {volume} {3}},\ \bibinfo {pages} {659} (\bibinfo {year} {2016})}\BibitemShut {NoStop}%
\bibitem [{\citenamefont {Ding}\ \emph {et~al.}(2020)\citenamefont {Ding}, \citenamefont {Wu}, \citenamefont {Mazidi}, \citenamefont {Zhang},\ and\ \citenamefont {Lew}}]{ding2020single}%
  \BibitemOpen
  \bibfield  {author} {\bibinfo {author} {\bibfnamefont {T.}~\bibnamefont {Ding}}, \bibinfo {author} {\bibfnamefont {T.}~\bibnamefont {Wu}}, \bibinfo {author} {\bibfnamefont {H.}~\bibnamefont {Mazidi}}, \bibinfo {author} {\bibfnamefont {O.}~\bibnamefont {Zhang}}, \ and\ \bibinfo {author} {\bibfnamefont {M.~D.}\ \bibnamefont {Lew}},\ }\href@noop {} {\bibfield  {journal} {\bibinfo  {journal} {Optica}\ }\textbf {\bibinfo {volume} {7}},\ \bibinfo {pages} {602} (\bibinfo {year} {2020})}\BibitemShut {NoStop}%
\bibitem [{\citenamefont {Rimoli}\ \emph {et~al.}(2022)\citenamefont {Rimoli}, \citenamefont {Valades-Cruz}, \citenamefont {Curcio}, \citenamefont {Mavrakis},\ and\ \citenamefont {Brasselet}}]{rimoli20224polar}%
  \BibitemOpen
  \bibfield  {author} {\bibinfo {author} {\bibfnamefont {C.~V.}\ \bibnamefont {Rimoli}}, \bibinfo {author} {\bibfnamefont {C.~A.}\ \bibnamefont {Valades-Cruz}}, \bibinfo {author} {\bibfnamefont {V.}~\bibnamefont {Curcio}}, \bibinfo {author} {\bibfnamefont {M.}~\bibnamefont {Mavrakis}}, \ and\ \bibinfo {author} {\bibfnamefont {S.}~\bibnamefont {Brasselet}},\ }\href@noop {} {\bibfield  {journal} {\bibinfo  {journal} {Nat. Commun.}\ }\textbf {\bibinfo {volume} {13}},\ \bibinfo {pages} {301} (\bibinfo {year} {2022})}\BibitemShut {NoStop}%
\bibitem [{\citenamefont {Zhang}\ \emph {et~al.}(2023)\citenamefont {Zhang}, \citenamefont {Guo}, \citenamefont {He}, \citenamefont {Wu}, \citenamefont {Vahey},\ and\ \citenamefont {Lew}}]{zhang2023six}%
  \BibitemOpen
  \bibfield  {author} {\bibinfo {author} {\bibfnamefont {O.}~\bibnamefont {Zhang}}, \bibinfo {author} {\bibfnamefont {Z.}~\bibnamefont {Guo}}, \bibinfo {author} {\bibfnamefont {Y.}~\bibnamefont {He}}, \bibinfo {author} {\bibfnamefont {T.}~\bibnamefont {Wu}}, \bibinfo {author} {\bibfnamefont {M.~D.}\ \bibnamefont {Vahey}}, \ and\ \bibinfo {author} {\bibfnamefont {M.~D.}\ \bibnamefont {Lew}},\ }\href@noop {} {\bibfield  {journal} {\bibinfo  {journal} {Nat. Photon.}\ }\textbf {\bibinfo {volume} {17}},\ \bibinfo {pages} {179} (\bibinfo {year} {2023})}\BibitemShut {NoStop}%
\bibitem [{\citenamefont {Curcio}\ \emph {et~al.}(2020)\citenamefont {Curcio}, \citenamefont {Alem{\'a}n-Casta{\~n}eda}, \citenamefont {Brown}, \citenamefont {Brasselet},\ and\ \citenamefont {Alonso}}]{curcio2020birefringent}%
  \BibitemOpen
  \bibfield  {author} {\bibinfo {author} {\bibfnamefont {V.}~\bibnamefont {Curcio}}, \bibinfo {author} {\bibfnamefont {L.~A.}\ \bibnamefont {Alem{\'a}n-Casta{\~n}eda}}, \bibinfo {author} {\bibfnamefont {T.~G.}\ \bibnamefont {Brown}}, \bibinfo {author} {\bibfnamefont {S.}~\bibnamefont {Brasselet}}, \ and\ \bibinfo {author} {\bibfnamefont {M.~A.}\ \bibnamefont {Alonso}},\ }\href@noop {} {\bibfield  {journal} {\bibinfo  {journal} {Nat. Commun.}\ }\textbf {\bibinfo {volume} {11}},\ \bibinfo {pages} {5307} (\bibinfo {year} {2020})}\BibitemShut {NoStop}%
\bibitem [{\citenamefont {Ding}\ and\ \citenamefont {Lew}(2021)}]{ding2021single}%
  \BibitemOpen
  \bibfield  {author} {\bibinfo {author} {\bibfnamefont {T.}~\bibnamefont {Ding}}\ and\ \bibinfo {author} {\bibfnamefont {M.~D.}\ \bibnamefont {Lew}},\ }\href@noop {} {\bibfield  {journal} {\bibinfo  {journal} {J. Phys. Chem. B}\ }\textbf {\bibinfo {volume} {125}},\ \bibinfo {pages} {12718} (\bibinfo {year} {2021})}\BibitemShut {NoStop}%
\bibitem [{\citenamefont {Hulleman}\ \emph {et~al.}(2021)\citenamefont {Hulleman}, \citenamefont {Thorsen}, \citenamefont {Kim}, \citenamefont {Dekker}, \citenamefont {Stallinga},\ and\ \citenamefont {Rieger}}]{hulleman2021simultaneous}%
  \BibitemOpen
  \bibfield  {author} {\bibinfo {author} {\bibfnamefont {C.~N.}\ \bibnamefont {Hulleman}}, \bibinfo {author} {\bibfnamefont {R.~{\O}.}\ \bibnamefont {Thorsen}}, \bibinfo {author} {\bibfnamefont {E.}~\bibnamefont {Kim}}, \bibinfo {author} {\bibfnamefont {C.}~\bibnamefont {Dekker}}, \bibinfo {author} {\bibfnamefont {S.}~\bibnamefont {Stallinga}}, \ and\ \bibinfo {author} {\bibfnamefont {B.}~\bibnamefont {Rieger}},\ }\href@noop {} {\bibfield  {journal} {\bibinfo  {journal} {Nat. Commun.}\ }\textbf {\bibinfo {volume} {12}},\ \bibinfo {pages} {5934} (\bibinfo {year} {2021})}\BibitemShut {NoStop}%
\bibitem [{\citenamefont {Empedocles}\ \emph {et~al.}(1999{\natexlab{a}})\citenamefont {Empedocles}, \citenamefont {Neuhauser},\ and\ \citenamefont {Bawendi}}]{empedocles1999three}%
  \BibitemOpen
  \bibfield  {author} {\bibinfo {author} {\bibfnamefont {S.~A.}\ \bibnamefont {Empedocles}}, \bibinfo {author} {\bibfnamefont {R.}~\bibnamefont {Neuhauser}}, \ and\ \bibinfo {author} {\bibfnamefont {M.~G.}\ \bibnamefont {Bawendi}},\ }\href@noop {} {\bibfield  {journal} {\bibinfo  {journal} {Nature}\ }\textbf {\bibinfo {volume} {399}},\ \bibinfo {pages} {126} (\bibinfo {year} {1999}{\natexlab{a}})}\BibitemShut {NoStop}%
\bibitem [{\citenamefont {Empedocles}\ \emph {et~al.}(1999{\natexlab{b}})\citenamefont {Empedocles}, \citenamefont {Neuhauser}, \citenamefont {Shimizu},\ and\ \citenamefont {Bawendi}}]{empedocles1999photoluminescence}%
  \BibitemOpen
  \bibfield  {author} {\bibinfo {author} {\bibfnamefont {S.~A.}\ \bibnamefont {Empedocles}}, \bibinfo {author} {\bibfnamefont {R.}~\bibnamefont {Neuhauser}}, \bibinfo {author} {\bibfnamefont {K.}~\bibnamefont {Shimizu}}, \ and\ \bibinfo {author} {\bibfnamefont {M.~G.}\ \bibnamefont {Bawendi}},\ }\href@noop {} {\bibfield  {journal} {\bibinfo  {journal} {Adv. Mater.}\ }\textbf {\bibinfo {volume} {11}},\ \bibinfo {pages} {1243} (\bibinfo {year} {1999}{\natexlab{b}})}\BibitemShut {NoStop}%
\bibitem [{\citenamefont {Ma}\ \emph {et~al.}(2018)\citenamefont {Ma}, \citenamefont {Diroll}, \citenamefont {Cho}, \citenamefont {Fedin}, \citenamefont {Schaller}, \citenamefont {Talapin},\ and\ \citenamefont {Wiederrecht}}]{ma2018anisotropic}%
  \BibitemOpen
  \bibfield  {author} {\bibinfo {author} {\bibfnamefont {X.}~\bibnamefont {Ma}}, \bibinfo {author} {\bibfnamefont {B.~T.}\ \bibnamefont {Diroll}}, \bibinfo {author} {\bibfnamefont {W.}~\bibnamefont {Cho}}, \bibinfo {author} {\bibfnamefont {I.}~\bibnamefont {Fedin}}, \bibinfo {author} {\bibfnamefont {R.~D.}\ \bibnamefont {Schaller}}, \bibinfo {author} {\bibfnamefont {D.~V.}\ \bibnamefont {Talapin}}, \ and\ \bibinfo {author} {\bibfnamefont {G.~P.}\ \bibnamefont {Wiederrecht}},\ }\href@noop {} {\bibfield  {journal} {\bibinfo  {journal} {Nano Lett.}\ }\textbf {\bibinfo {volume} {18}},\ \bibinfo {pages} {4647} (\bibinfo {year} {2018})}\BibitemShut {NoStop}%
\bibitem [{\citenamefont {Chizhik}\ \emph {et~al.}(2011)\citenamefont {Chizhik}, \citenamefont {Chizhik}, \citenamefont {Khoptyar}, \citenamefont {B\"ar},\ and\ \citenamefont {Meixner}}]{chizhik2011excitation}%
  \BibitemOpen
  \bibfield  {author} {\bibinfo {author} {\bibfnamefont {A.~I.}\ \bibnamefont {Chizhik}}, \bibinfo {author} {\bibfnamefont {A.~M.}\ \bibnamefont {Chizhik}}, \bibinfo {author} {\bibfnamefont {D.}~\bibnamefont {Khoptyar}}, \bibinfo {author} {\bibfnamefont {S.}~\bibnamefont {B\"ar}}, \ and\ \bibinfo {author} {\bibfnamefont {A.~J.}\ \bibnamefont {Meixner}},\ }\href@noop {} {\bibfield  {journal} {\bibinfo  {journal} {Nano Lett.}\ }\textbf {\bibinfo {volume} {11}},\ \bibinfo {pages} {1131} (\bibinfo {year} {2011})}\BibitemShut {NoStop}%
\bibitem [{\citenamefont {Ghosh}\ \emph {et~al.}(2019)\citenamefont {Ghosh}, \citenamefont {Chizhik}, \citenamefont {Yang}, \citenamefont {Karedla}, \citenamefont {Gregor}, \citenamefont {Oron}, \citenamefont {Weiss}, \citenamefont {Enderlein},\ and\ \citenamefont {Chizhik}}]{ghosh2019excitation}%
  \BibitemOpen
  \bibfield  {author} {\bibinfo {author} {\bibfnamefont {S.}~\bibnamefont {Ghosh}}, \bibinfo {author} {\bibfnamefont {A.~M.}\ \bibnamefont {Chizhik}}, \bibinfo {author} {\bibfnamefont {G.}~\bibnamefont {Yang}}, \bibinfo {author} {\bibfnamefont {N.}~\bibnamefont {Karedla}}, \bibinfo {author} {\bibfnamefont {I.}~\bibnamefont {Gregor}}, \bibinfo {author} {\bibfnamefont {D.}~\bibnamefont {Oron}}, \bibinfo {author} {\bibfnamefont {S.}~\bibnamefont {Weiss}}, \bibinfo {author} {\bibfnamefont {J.}~\bibnamefont {Enderlein}}, \ and\ \bibinfo {author} {\bibfnamefont {A.~I.}\ \bibnamefont {Chizhik}},\ }\href@noop {} {\bibfield  {journal} {\bibinfo  {journal} {Nano Lett.}\ }\textbf {\bibinfo {volume} {19}},\ \bibinfo {pages} {1695} (\bibinfo {year} {2019})}\BibitemShut {NoStop}%
\bibitem [{\citenamefont {Zhang}\ \emph {et~al.}(2022)\citenamefont {Zhang}, \citenamefont {Zhou}, \citenamefont {Lu}, \citenamefont {Wu},\ and\ \citenamefont {Lew}}]{zhang2022resolving}%
  \BibitemOpen
  \bibfield  {author} {\bibinfo {author} {\bibfnamefont {O.}~\bibnamefont {Zhang}}, \bibinfo {author} {\bibfnamefont {W.}~\bibnamefont {Zhou}}, \bibinfo {author} {\bibfnamefont {J.}~\bibnamefont {Lu}}, \bibinfo {author} {\bibfnamefont {T.}~\bibnamefont {Wu}}, \ and\ \bibinfo {author} {\bibfnamefont {M.~D.}\ \bibnamefont {Lew}},\ }\href@noop {} {\bibfield  {journal} {\bibinfo  {journal} {Nano Lett.}\ }\textbf {\bibinfo {volume} {22}},\ \bibinfo {pages} {1024} (\bibinfo {year} {2022})}\BibitemShut {NoStop}%
\bibitem [{\citenamefont {Rotenberg}\ and\ \citenamefont {Kuipers}(2014)}]{rotenberg2014mapping}%
  \BibitemOpen
  \bibfield  {author} {\bibinfo {author} {\bibfnamefont {N.}~\bibnamefont {Rotenberg}}\ and\ \bibinfo {author} {\bibfnamefont {L.}~\bibnamefont {Kuipers}},\ }\href@noop {} {\bibfield  {journal} {\bibinfo  {journal} {Nat. Photon.}\ }\textbf {\bibinfo {volume} {8}},\ \bibinfo {pages} {919} (\bibinfo {year} {2014})}\BibitemShut {NoStop}%
\bibitem [{\citenamefont {Cang}\ \emph {et~al.}(2011)\citenamefont {Cang}, \citenamefont {Labno}, \citenamefont {Lu}, \citenamefont {Yin}, \citenamefont {Liu}, \citenamefont {Gladden}, \citenamefont {Liu},\ and\ \citenamefont {Zhang}}]{cang2011probing}%
  \BibitemOpen
  \bibfield  {author} {\bibinfo {author} {\bibfnamefont {H.}~\bibnamefont {Cang}}, \bibinfo {author} {\bibfnamefont {A.}~\bibnamefont {Labno}}, \bibinfo {author} {\bibfnamefont {C.}~\bibnamefont {Lu}}, \bibinfo {author} {\bibfnamefont {X.}~\bibnamefont {Yin}}, \bibinfo {author} {\bibfnamefont {M.}~\bibnamefont {Liu}}, \bibinfo {author} {\bibfnamefont {C.}~\bibnamefont {Gladden}}, \bibinfo {author} {\bibfnamefont {Y.}~\bibnamefont {Liu}}, \ and\ \bibinfo {author} {\bibfnamefont {X.}~\bibnamefont {Zhang}},\ }\href@noop {} {\bibfield  {journal} {\bibinfo  {journal} {Nature}\ }\textbf {\bibinfo {volume} {469}},\ \bibinfo {pages} {385} (\bibinfo {year} {2011})}\BibitemShut {NoStop}%
\bibitem [{\citenamefont {Singh}\ \emph {et~al.}(2014)\citenamefont {Singh}, \citenamefont {Calbris},\ and\ \citenamefont {Van~Hulst}}]{singh2014vectorial}%
  \BibitemOpen
  \bibfield  {author} {\bibinfo {author} {\bibfnamefont {A.}~\bibnamefont {Singh}}, \bibinfo {author} {\bibfnamefont {G.}~\bibnamefont {Calbris}}, \ and\ \bibinfo {author} {\bibfnamefont {N.~F.}\ \bibnamefont {Van~Hulst}},\ }\href@noop {} {\bibfield  {journal} {\bibinfo  {journal} {Nano Lett.}\ }\textbf {\bibinfo {volume} {14}},\ \bibinfo {pages} {4715} (\bibinfo {year} {2014})}\BibitemShut {NoStop}%
\bibitem [{\citenamefont {Steuwe}\ \emph {et~al.}(2015)\citenamefont {Steuwe}, \citenamefont {Erdelyi}, \citenamefont {Szekeres}, \citenamefont {Csete}, \citenamefont {Baumberg}, \citenamefont {Mahajan},\ and\ \citenamefont {Kaminski}}]{steuwe2015visualizing}%
  \BibitemOpen
  \bibfield  {author} {\bibinfo {author} {\bibfnamefont {C.}~\bibnamefont {Steuwe}}, \bibinfo {author} {\bibfnamefont {M.}~\bibnamefont {Erdelyi}}, \bibinfo {author} {\bibfnamefont {G.}~\bibnamefont {Szekeres}}, \bibinfo {author} {\bibfnamefont {M.}~\bibnamefont {Csete}}, \bibinfo {author} {\bibfnamefont {J.~J.}\ \bibnamefont {Baumberg}}, \bibinfo {author} {\bibfnamefont {S.}~\bibnamefont {Mahajan}}, \ and\ \bibinfo {author} {\bibfnamefont {C.~F.}\ \bibnamefont {Kaminski}},\ }\href@noop {} {\bibfield  {journal} {\bibinfo  {journal} {Nano Lett.}\ }\textbf {\bibinfo {volume} {15}},\ \bibinfo {pages} {3217} (\bibinfo {year} {2015})}\BibitemShut {NoStop}%
\bibitem [{\citenamefont {Mack}\ \emph {et~al.}(2017)\citenamefont {Mack}, \citenamefont {Cort{\'e}s}, \citenamefont {Giannini}, \citenamefont {T{\"o}r{\"o}k}, \citenamefont {Roschuk},\ and\ \citenamefont {Maier}}]{mack2017decoupling}%
  \BibitemOpen
  \bibfield  {author} {\bibinfo {author} {\bibfnamefont {D.~L.}\ \bibnamefont {Mack}}, \bibinfo {author} {\bibfnamefont {E.}~\bibnamefont {Cort{\'e}s}}, \bibinfo {author} {\bibfnamefont {V.}~\bibnamefont {Giannini}}, \bibinfo {author} {\bibfnamefont {P.}~\bibnamefont {T{\"o}r{\"o}k}}, \bibinfo {author} {\bibfnamefont {T.}~\bibnamefont {Roschuk}}, \ and\ \bibinfo {author} {\bibfnamefont {S.~A.}\ \bibnamefont {Maier}},\ }\href@noop {} {\bibfield  {journal} {\bibinfo  {journal} {Nat. Commun.}\ }\textbf {\bibinfo {volume} {8}},\ \bibinfo {pages} {14513} (\bibinfo {year} {2017})}\BibitemShut {NoStop}%
\bibitem [{\citenamefont {Nov{\'a}k}\ \emph {et~al.}(2025)\citenamefont {Nov{\'a}k}, \citenamefont {B{\'\i}r{\'o}}, \citenamefont {Ferenc}, \citenamefont {Ungor}, \citenamefont {Czvik}, \citenamefont {De{\'a}k}, \citenamefont {Janov{\'a}k},\ and\ \citenamefont {Erd{\'e}lyi}}]{novak2025using}%
  \BibitemOpen
  \bibfield  {author} {\bibinfo {author} {\bibfnamefont {T.}~\bibnamefont {Nov{\'a}k}}, \bibinfo {author} {\bibfnamefont {P.}~\bibnamefont {B{\'\i}r{\'o}}}, \bibinfo {author} {\bibfnamefont {G.}~\bibnamefont {Ferenc}}, \bibinfo {author} {\bibfnamefont {D.}~\bibnamefont {Ungor}}, \bibinfo {author} {\bibfnamefont {E.}~\bibnamefont {Czvik}}, \bibinfo {author} {\bibfnamefont {{\'A}.}~\bibnamefont {De{\'a}k}}, \bibinfo {author} {\bibfnamefont {L.}~\bibnamefont {Janov{\'a}k}}, \ and\ \bibinfo {author} {\bibfnamefont {M.}~\bibnamefont {Erd{\'e}lyi}},\ }\href@noop {} {\bibfield  {journal} {\bibinfo  {journal} {Opt. Commun.}\ }\textbf {\bibinfo {volume} {574}},\ \bibinfo {pages} {131147} (\bibinfo {year} {2025})}\BibitemShut {NoStop}%
\bibitem [{\citenamefont {Wolf}(1959)}]{wolf1959electromagnetic}%
  \BibitemOpen
  \bibfield  {author} {\bibinfo {author} {\bibfnamefont {E.}~\bibnamefont {Wolf}},\ }\href@noop {} {\bibfield  {journal} {\bibinfo  {journal} {Proc. R. Soc. Lond. Ser. A Math. Phys. Sci.}\ }\textbf {\bibinfo {volume} {253}},\ \bibinfo {pages} {349} (\bibinfo {year} {1959})}\BibitemShut {NoStop}%
\bibitem [{\citenamefont {Richards}\ and\ \citenamefont {Wolf}(1959)}]{richards1959electromagnetic}%
  \BibitemOpen
  \bibfield  {author} {\bibinfo {author} {\bibfnamefont {B.}~\bibnamefont {Richards}}\ and\ \bibinfo {author} {\bibfnamefont {E.}~\bibnamefont {Wolf}},\ }\href@noop {} {\bibfield  {journal} {\bibinfo  {journal} {Proc. R. Soc. Lond. Ser. A Math. Phys. Sci.}\ }\textbf {\bibinfo {volume} {253}},\ \bibinfo {pages} {358} (\bibinfo {year} {1959})}\BibitemShut {NoStop}%
\bibitem [{\citenamefont {Fazel}\ \emph {et~al.}(2024)\citenamefont {Fazel}, \citenamefont {Grussmayer}, \citenamefont {Ferdman}, \citenamefont {Radenovic}, \citenamefont {Shechtman}, \citenamefont {Enderlein},\ and\ \citenamefont {Press{\'e}}}]{fazel2024fluorescence}%
  \BibitemOpen
  \bibfield  {author} {\bibinfo {author} {\bibfnamefont {M.}~\bibnamefont {Fazel}}, \bibinfo {author} {\bibfnamefont {K.~S.}\ \bibnamefont {Grussmayer}}, \bibinfo {author} {\bibfnamefont {B.}~\bibnamefont {Ferdman}}, \bibinfo {author} {\bibfnamefont {A.}~\bibnamefont {Radenovic}}, \bibinfo {author} {\bibfnamefont {Y.}~\bibnamefont {Shechtman}}, \bibinfo {author} {\bibfnamefont {J.}~\bibnamefont {Enderlein}}, \ and\ \bibinfo {author} {\bibfnamefont {S.}~\bibnamefont {Press{\'e}}},\ }\href@noop {} {\bibfield  {journal} {\bibinfo  {journal} {Rev. Mod. Phys.}\ }\textbf {\bibinfo {volume} {96}},\ \bibinfo {pages} {025003} (\bibinfo {year} {2024})}\BibitemShut {NoStop}%
\bibitem [{\citenamefont {Leutenegger}\ \emph {et~al.}(2006)\citenamefont {Leutenegger}, \citenamefont {Rao}, \citenamefont {Leitgeb},\ and\ \citenamefont {Lasser}}]{leutenegger2006fast}%
  \BibitemOpen
  \bibfield  {author} {\bibinfo {author} {\bibfnamefont {M.}~\bibnamefont {Leutenegger}}, \bibinfo {author} {\bibfnamefont {R.}~\bibnamefont {Rao}}, \bibinfo {author} {\bibfnamefont {R.~A.}\ \bibnamefont {Leitgeb}}, \ and\ \bibinfo {author} {\bibfnamefont {T.}~\bibnamefont {Lasser}},\ }\href@noop {} {\bibfield  {journal} {\bibinfo  {journal} {Opt. Express}\ }\textbf {\bibinfo {volume} {14}},\ \bibinfo {pages} {11277} (\bibinfo {year} {2006})}\BibitemShut {NoStop}%
\bibitem [{\citenamefont {James}\ \emph {et~al.}(2019)\citenamefont {James}, \citenamefont {Unni}, \citenamefont {Taleb}, \citenamefont {Chapel}, \citenamefont {Kalarikkal}, \citenamefont {Varghese}, \citenamefont {Vignaud},\ and\ \citenamefont {Grohens}}]{james2019surface}%
  \BibitemOpen
  \bibfield  {author} {\bibinfo {author} {\bibfnamefont {J.}~\bibnamefont {James}}, \bibinfo {author} {\bibfnamefont {A.~B.}\ \bibnamefont {Unni}}, \bibinfo {author} {\bibfnamefont {K.}~\bibnamefont {Taleb}}, \bibinfo {author} {\bibfnamefont {J.-P.}\ \bibnamefont {Chapel}}, \bibinfo {author} {\bibfnamefont {N.}~\bibnamefont {Kalarikkal}}, \bibinfo {author} {\bibfnamefont {S.}~\bibnamefont {Varghese}}, \bibinfo {author} {\bibfnamefont {G.}~\bibnamefont {Vignaud}}, \ and\ \bibinfo {author} {\bibfnamefont {Y.}~\bibnamefont {Grohens}},\ }\href@noop {} {\bibfield  {journal} {\bibinfo  {journal} {Nano-Struct. Nano-Objects}\ }\textbf {\bibinfo {volume} {17}},\ \bibinfo {pages} {34} (\bibinfo {year} {2019})}\BibitemShut {NoStop}%
\bibitem [{\citenamefont {Vignaud}\ \emph {et~al.}(2014)\citenamefont {Vignaud}, \citenamefont {S.~Chebil}, \citenamefont {Bal}, \citenamefont {Delorme}, \citenamefont {Beuvier}, \citenamefont {Grohens},\ and\ \citenamefont {Gibaud}}]{vignaud2014densification}%
  \BibitemOpen
  \bibfield  {author} {\bibinfo {author} {\bibfnamefont {G.}~\bibnamefont {Vignaud}}, \bibinfo {author} {\bibfnamefont {M.}~\bibnamefont {S.~Chebil}}, \bibinfo {author} {\bibfnamefont {J.}~\bibnamefont {Bal}}, \bibinfo {author} {\bibfnamefont {N.}~\bibnamefont {Delorme}}, \bibinfo {author} {\bibfnamefont {T.}~\bibnamefont {Beuvier}}, \bibinfo {author} {\bibfnamefont {Y.}~\bibnamefont {Grohens}}, \ and\ \bibinfo {author} {\bibfnamefont {A.}~\bibnamefont {Gibaud}},\ }\href@noop {} {\bibfield  {journal} {\bibinfo  {journal} {Langmuir}\ }\textbf {\bibinfo {volume} {30}},\ \bibinfo {pages} {11599} (\bibinfo {year} {2014})}\BibitemShut {NoStop}%
\bibitem [{\citenamefont {Lewis}(1995)}]{lewis1995fast}%
  \BibitemOpen
  \bibfield  {author} {\bibinfo {author} {\bibfnamefont {J.~P.}\ \bibnamefont {Lewis}},\ }in\ \href@noop {} {\emph {\bibinfo {booktitle} {Vision interface}}},\ Vol.~\bibinfo {volume} {10}\ (\bibinfo {year} {1995})\ pp.\ \bibinfo {pages} {120--123}\BibitemShut {NoStop}%
\bibitem [{\citenamefont {Radmacher}(2025)}]{radmacher2025fluorescence}%
  \BibitemOpen
  \bibfield  {author} {\bibinfo {author} {\bibfnamefont {N.~J.}\ \bibnamefont {Radmacher}},\ }\emph {\bibinfo {title} {Fluorescence-lifetime image scanning microscopy}},\ \href {\doibase 10.53846/goediss-11145} {\bibinfo {type} {Phd thesis}},\ \bibinfo  {school} {Georg-August-Universit{\"a}t G{\"o}ttingen} (\bibinfo {year} {2025})\BibitemShut {NoStop}%
\bibitem [{\citenamefont {Bliokh}\ \emph {et~al.}(2015)\citenamefont {Bliokh}, \citenamefont {Rodr{\'\i}guez-Fortu{\~n}o}, \citenamefont {Nori},\ and\ \citenamefont {Zayats}}]{bliokh2015spin}%
  \BibitemOpen
  \bibfield  {author} {\bibinfo {author} {\bibfnamefont {K.~Y.}\ \bibnamefont {Bliokh}}, \bibinfo {author} {\bibfnamefont {F.~J.}\ \bibnamefont {Rodr{\'\i}guez-Fortu{\~n}o}}, \bibinfo {author} {\bibfnamefont {F.}~\bibnamefont {Nori}}, \ and\ \bibinfo {author} {\bibfnamefont {A.~V.}\ \bibnamefont {Zayats}},\ }\href@noop {} {\bibfield  {journal} {\bibinfo  {journal} {Nat. Photon.}\ }\textbf {\bibinfo {volume} {9}},\ \bibinfo {pages} {796} (\bibinfo {year} {2015})}\BibitemShut {NoStop}%
\bibitem [{\citenamefont {B{\"o}rner}\ \emph {et~al.}(2016)\citenamefont {B{\"o}rner}, \citenamefont {Ehrlich}, \citenamefont {Hohlbein},\ and\ \citenamefont {H{\"u}bner}}]{borner2016single}%
  \BibitemOpen
  \bibfield  {author} {\bibinfo {author} {\bibfnamefont {R.}~\bibnamefont {B{\"o}rner}}, \bibinfo {author} {\bibfnamefont {N.}~\bibnamefont {Ehrlich}}, \bibinfo {author} {\bibfnamefont {J.}~\bibnamefont {Hohlbein}}, \ and\ \bibinfo {author} {\bibfnamefont {C.~G.}\ \bibnamefont {H{\"u}bner}},\ }\href@noop {} {\bibfield  {journal} {\bibinfo  {journal} {J. Fluoresc.}\ }\textbf {\bibinfo {volume} {26}},\ \bibinfo {pages} {963} (\bibinfo {year} {2016})}\BibitemShut {NoStop}%
\bibitem [{\citenamefont {Adamczyk}\ \emph {et~al.}(2022)\citenamefont {Adamczyk}, \citenamefont {Huijben}, \citenamefont {Sison}, \citenamefont {Di~Luca}, \citenamefont {Chiarelli}, \citenamefont {Vanni}, \citenamefont {Brasselet}, \citenamefont {Mortensen}, \citenamefont {Stefani}, \citenamefont {Pilo-Pais},\ and\ \citenamefont {Acuna}}]{adamczyk2022dna}%
  \BibitemOpen
  \bibfield  {author} {\bibinfo {author} {\bibfnamefont {A.~K.}\ \bibnamefont {Adamczyk}}, \bibinfo {author} {\bibfnamefont {T.~A.}\ \bibnamefont {Huijben}}, \bibinfo {author} {\bibfnamefont {M.}~\bibnamefont {Sison}}, \bibinfo {author} {\bibfnamefont {A.}~\bibnamefont {Di~Luca}}, \bibinfo {author} {\bibfnamefont {G.}~\bibnamefont {Chiarelli}}, \bibinfo {author} {\bibfnamefont {S.}~\bibnamefont {Vanni}}, \bibinfo {author} {\bibfnamefont {S.}~\bibnamefont {Brasselet}}, \bibinfo {author} {\bibfnamefont {K.~I.}\ \bibnamefont {Mortensen}}, \bibinfo {author} {\bibfnamefont {F.~D.}\ \bibnamefont {Stefani}}, \bibinfo {author} {\bibfnamefont {M.}~\bibnamefont {Pilo-Pais}}, \ and\ \bibinfo {author} {\bibfnamefont {G.~P.}\ \bibnamefont {Acuna}},\ }\href@noop {} {\bibfield  {journal} {\bibinfo  {journal} {ACS Nano}\ }\textbf {\bibinfo {volume} {16}},\ \bibinfo {pages} {16924} (\bibinfo {year} {2022})}\BibitemShut {NoStop}%
\bibitem [{\citenamefont {Pressly}\ \emph {et~al.}(2019)\citenamefont {Pressly}, \citenamefont {Riggleman},\ and\ \citenamefont {Winey}}]{pressly2019increased}%
  \BibitemOpen
  \bibfield  {author} {\bibinfo {author} {\bibfnamefont {J.~F.}\ \bibnamefont {Pressly}}, \bibinfo {author} {\bibfnamefont {R.~A.}\ \bibnamefont {Riggleman}}, \ and\ \bibinfo {author} {\bibfnamefont {K.~I.}\ \bibnamefont {Winey}},\ }\href@noop {} {\bibfield  {journal} {\bibinfo  {journal} {Macromol.}\ }\textbf {\bibinfo {volume} {52}},\ \bibinfo {pages} {6116} (\bibinfo {year} {2019})}\BibitemShut {NoStop}%
\end{thebibliography}
%

\end{document}